\documentclass[conference]{IEEEtran}
\IEEEoverridecommandlockouts

\usepackage{algorithmic}
\usepackage{textcomp}

\usepackage{amsmath,amsfonts,amssymb,bm}
\usepackage{graphicx}
\usepackage{xcolor}
\usepackage{cite}
\usepackage{booktabs}
\usepackage{tikz}
\usepackage{pgfplots}
\usepackage{multirow}
\usepackage[acronym,shortcuts]{glossaries}
\usepackage[caption=false,font=small,labelfont=rm,textfont=rm]{subfig}
\usepackage{amsthm}

\pgfplotsset{compat=1.18}
\usepackage{comment}

\def\BibTeX{{\rm B\kern-.05em{\sc i\kern-.025em b}\kern-.08em
    T\kern-.1667em\lower.7ex\hbox{E}\kern-.125emX}}

\newcommand{\E}{\mathbb{E}}

\newacronym{DL}{DL}{deep learning}
\newacronym{NN}{NN}{neural network}
\newacronym{ReLU}{ReLU}{rectified linear unit}
\newacronym{BP}{BP}{belief propagation}
\newacronym{MSE}{MSE}{mean squared error}
\newacronym{MRC}{MRC}{maximal ratio combining}
\newacronym{CRC}{CRC}{cyclic redundancy check}
\newacronym{SISO}{SISO}{soft-input soft-output}
\newacronym{AWGN}{AWGN}{additive white Gaussian noise}
\newacronym{BPSK}{BPSK}{binary phase-shift keying}
\newacronym{C2C}{C2C}{control-to-control}
\newacronym{CNN}{CNN}{convolutional neural network}
\newacronym{GLRT}{GLRT}{generalized likelihood ratio test}
\newacronym{LDPC}{LDPC}{low-density parity-check}
\newacronym{LLR}{LLR}{log-likelihood ratio}
\newacronym{NMS}{NMS}{normalized min-sum}
\newacronym{PLR}{PLR}{packet loss rate}
\newacronym{SIC}{SIC}{successive interference cancellation}
\newacronym{PER}{PER}{packet error rate}
\newacronym{UPER}{UPER}{undetected packet error rate}
\newacronym{ROC}{ROC}{receiver operating characteristic}
\newacronym{FLOP}{FLOP}{floating point operation}

\begin{document}

\title{Short-Boundary Signaling for Asynchronous Grant-Free Control-to-Control Communications}

\author{Massimo Battaglioni, Dania De Crescenzo, Edoardo Carnevali, and Enrico Paolini%
\thanks{M. Battaglioni and E. Carnevali are with the Department of Information Engineering, Universit\`a Politecnica delle Marche, 60131 Ancona, Italy (e-mail: m.battaglioni@univpm.it; edoardo.carnevali@pm.univpm.it). D. De Crescenzo and E. Paolini are with CNIT/WiLab, DEI, University of Bologna, 47522 Cesena, Italy (e-mail: dania.decrescenzo2@unibo.it; e.paolini@unibo.it). Corresponding author: M. Battaglioni.}}

\maketitle

\begin{abstract}
In asynchronous grant-free access, receivers must detect variable-length transmissions and identify their start and end.  Existing decoder-aided receivers use start and tail sequences which are as long as one codeword, causing substantial overhead for short commands. This paper treats the boundary (i.e., start and tail sequences) length as an independent design parameter and proposes a convolutional neural network that detects both boundaries from the received waveform and the decoder soft output. Its waveform input size decreases with the boundary length, while the decoder input keeps the codeword length, so the boundary can be shortened without altering the decoder. Analysis and simulations quantify the resulting tradeoff among payload efficiency, collision exposure, and boundary detection. The results show that the proposed receiver can exploit very short boundaries: in the considered setting, the shortest tested boundary, $L_{\mathrm{b}}=16$, yields the highest effective payload efficiency at all tested loads, whereas a correlation-based benchmark with the same sequences performs best with $L_{\mathrm{b}}=64$ at moderate and high loads.
\end{abstract}

\begin{IEEEkeywords}
Asynchronous random access, control-to-control communications, deep learning, grant-free access.
\end{IEEEkeywords}

\glsresetall

\section{Introduction}

Distributed motion-control systems rely on direct communication among autonomous controllers in modular production lines, cooperative robotics, and reconfigurable industrial machinery~\cite{Aceto2019,Cuozzo2025}. Centralized scheduling may require continuous topology knowledge and introduce request-grant delays or communication bottlenecks~\cite{Chang2021,Gao2021}. Grant-free \ac{C2C} communications instead allow a controller to access the shared channel as soon as a command is generated.

Without access grants or a common time reference, packet acquisition and framing must be resolved before the receiver can reconstruct and cancel a transmission. In the considered architecture, command-length information is unavailable during acquisition, and the receiver must infer both packet boundaries directly from the received signal. Replica-based access and \ac{SIC} can resolve many collisions~\cite{DeGaudenzi2014,Paolini2015}, but only after the individual waveforms and their boundaries have been identified. Conventional synchronization commonly relies on correlation or a \ac{GLRT} applied to a known sequence~\cite{Chiani2006}. In the considered architecture, the boundary length affects both acquisition reliability and the duration and collision exposure of every transmitted replica. Deep-learning-based methods have also been proposed for activity, preamble, and replica detection in grant-free access~\cite{DeSouza2023,Khan2024,DeCrescenzo2025}. These methods are particularly suited to observations containing superposed codewords, boundary sequences, and interfering transmissions.

For variable-length commands, start detection might not be sufficient, since the receiver must also locate the packet termination before  reconstruction and interference cancellation can be completed. The receivers in~\cite{BattaglioniCSCN2025,BattaglioniArxiv2026} address this problem through command units comprising a start sequence, a variable number of \ac{LDPC} codewords, and a tail sequence. A single \ac{CNN} detects both boundaries and exploits the decoder soft trajectory to distinguish a tail from an additional codeword. Their main limitation is the architectural constraint that the start sequence, tail sequence, and one codeword must share the same length. This couples framing overhead and collision exposure to the code design, preventing the boundary duration from being optimized independently. For short commands, this coupling is particularly costly.

Building on~\cite{BattaglioniCSCN2025,BattaglioniArxiv2026}, this paper removes the  constraint on boundary and codeword lengths and introduces the former as an independent physical-layer design parameter. This  generalization changes several interacting properties of the receiver: reducing the boundary length lowers framing overhead, collision exposure, and \ac{CNN} inference cost, but simultaneously degrades boundary detection and the initial channel estimate. We characterize these competing effects analytically and modify the decoder-aided architecture of~\cite{BattaglioniArxiv2026} by assigning independent input dimensions to its two branches: the waveform-processing branch size scales with the boundary length, while the decoder-information branch retains the codeword length $n$. For the considered setting, the resulting design allows very short boundaries to be used without a prohibitive detection penalty, so that their reductions in framing overhead and collision exposure translate into end-to-end performance gains.

The remainder of the paper is organized as follows. Section~\ref{sec:sysmod} introduces the system model and performance metrics. Section~\ref{sec:shorten} analyzes the impact of boundary shortening. Section~\ref{sec:receiver} presents the proposed receiver, while Section~\ref{sec:numres} reports the numerical results. Finally, Section~\ref{sec:con} concludes the paper.

\section{System Model and Performance Metrics}
\label{sec:sysmod}
Throughout the paper, for integers $a\leq b$, $[a,b]\triangleq\{a,a+1,\ldots,b\}$ denotes the corresponding integer interval. 
\subsection{Transmission and Channel Model}
We consider distributed controllers exchanging variable-length commands over a shared wireless channel. Each transmission uses one antenna, and the receiver is equipped with $N_{\mathrm R}$ antennas. We adopt the asynchronous grant-free \ac{C2C} model of~\cite{BattaglioniArxiv2026}. At symbol time $t$, the received vector is
\begin{equation}
\mathbf r[t]=\sum_{j\in\mathcal A(t)}\mathbf h_j x_j[t-\tau_j]+\mathbf w[t],
\label{eq:received_signal}
\end{equation}
where $\mathcal A(t)$ is the set of replicas active at time $t$, $x_j[\cdot]$ is the transmitted symbol sequence of replica $j$, and $\tau_j\in\mathbb Z$ is its starting time. Asynchronism is represented by unknown integer symbol offsets. The channel vector $\mathbf h_j\in\mathbb C^{N_{\mathrm R}}$ is constant over each replica. We use the flat Rician fading, path-loss, and log-normal shadowing model of~\cite{BattaglioniArxiv2026}. Replicas from the same controller share the large-scale channel parameters, whereas their small-scale fading realizations are independent. The noise vectors are independent across symbol times, with $\mathbf w[t]\sim\mathcal{CN}(\mathbf 0,\sigma_w^2\mathbf I_{N_{\mathrm R}})$, where $\sigma_w^2$ is the noise variance per complex receive sample and $\mathbf I_{N_{\mathrm R}}$ is the identity matrix.

Command generations are modeled as a Poisson process of rate $\lambda$, where $\lambda$ is the average number of commands generated per symbol interval over all controllers. Each command is transmitted through $N_{\mathrm{rep}}$ replicas of the same command unit, placed within a local virtual frame. All controllers employ \ac{BPSK} and a common binary $(n,k)$ \ac{LDPC} code. A command containing $\iota\in[1,\iota_{\max}]$ codewords is transmitted as
\begin{equation}
\mathbf{x}=\left[\mathbf{x}_{\mathrm{s}},\mathbf{c}_1,\ldots,\mathbf{c}_{\iota},\mathbf{x}_{\mathrm{t}}\right],
\end{equation}
where $\mathbf{x}_{\mathrm{s}}\in\{-1,+1\}^{L_{\mathrm{s}}}$ and $\mathbf{x}_{\mathrm{t}}\in\{-1,+1\}^{L_{\mathrm{t}}}$ are the start and tail sequences, respectively, and each $\mathbf{c}_j\in\{-1,+1\}^{n}$ is a modulated codeword.  The command length is
\begin{equation}
L_{\iota}=\iota n+L_{\mathrm{s}}+L_{\mathrm{t}}.
\label{eq:length}
\end{equation}
Unlike~\cite{BattaglioniCSCN2025,BattaglioniArxiv2026}, we do not impose $L_{\mathrm{s}}=L_{\mathrm{t}}=n$. We focus on the symmetric design $L_{\mathrm{s}}=L_{\mathrm{t}}=L_{\mathrm{b}}$, which allows the effect of the boundary length to be studied independently. We use the length-$128$ start and tail sequences whose hexadecimal representations before \ac{BPSK} mapping are \texttt{1C71 B91B A9BA 8457 B4BC 5054 BFD0 5540} and \texttt{AA6C CB0C C243 AC5F 39DC 7AF4 640B 5D95}, respectively. The same pair is shared by all controllers and reused for all transmissions. For $L_{\mathrm b}<128$, the corresponding length-$L_{\mathrm b}$ boundaries are obtained by taking the first $L_{\mathrm b}$ bits of these parent sequences.

Each local virtual frame begins when a command is generated and contains $N_{\mathrm S}$ slots, each of length $L_{\max}=\iota_{\max}n+2L_{\mathrm b}$ symbols. Its duration is therefore $N_{\mathrm S}L_{\max}$ symbol intervals. The slots organize the replicas of each command locally: virtual frames from different controllers need not be aligned, i.e., controllers are not synchronized, and their replicas may overlap partially or completely. The receiver knows the code and the boundary sequences, but the replica starting times and the number of codewords in each command are unavailable before acquisition. The controller selects $N_{\mathrm{rep}}$ distinct slots uniformly at random and places one replica in each selected slot. For a command of length $L_{\iota}$, the starting offset within a selected slot is uniform over $[0,L_{\max}-L_{\iota}]$, independently across replicas. Slot selections are independent across controllers. 

\subsection{Receiver Operations}
The receiver searches the received waveform for a start sequence. A candidate starting time $\widehat{\tau}$ provides both a tentative payload alignment and a segment of known symbols for channel estimation. Given a channel estimate $\widehat{\mathbf h}$, the observations from the receive antennas are combined through \ac{MRC} as
\begin{equation}
\widehat{x}[t-\widehat{\tau}]=\frac{\widehat{\mathbf h}^{\mathsf H}\mathbf r[t]}{\|\widehat{\mathbf h}\|^2}.
\label{eq:mrc}
\end{equation}
The estimate obtained from the candidate start sequence is used to decode all codewords of the command; after successful decoding, it is refined over the complete reconstructed replica for cancellation, as described in Section~\ref{sec:shorten}.

Starting immediately after the detected start sequence, the receiver examines successive $n$-symbol blocks. At each possible codeword boundary, it must determine whether the next block contains another codeword or begins with the tail sequence. The \ac{LDPC} decoder processes the candidate block for $N_{\mathrm{it}}$ iterations and supplies bit \acp{LLR} and a convergence flag indicating whether the decoded word satisfies all parity checks. These outputs provide additional evidence for tail detection. For $L_{\mathrm b}<n$, a block beginning with a tail also contains samples beyond that replica's end; the waveform detector uses its first $L_{\mathrm b}$ samples, whereas the decoder still processes the complete $n$-symbol block.

Once a candidate command has passed decoding and boundary validation, all its replicas are cancelled. Their positions are assumed known once one replica is decoded. The channel of each replica is re-estimated over the complete reconstructed replica before subtraction. Detection and decoding are repeated on the residual signal through successive \ac{SIC} rounds, allowing replicas initially affected by interference to become recoverable. Section~\ref{sec:receiver} describes the proposed boundary detector used within this processing chain.

\subsection{Performance Metrics}

For a command containing $\iota$ codewords, the payload efficiency is defined as
\begin{equation}
\eta_{\mathrm{info}}(\iota,L_{\mathrm{b}})=\frac{\iota k}{\iota n+2L_{\mathrm{b}}}.
\label{eq:efficiency}\end{equation}
Let $I$ denote the random number of codewords per command (with realization $\iota$). The corresponding average payload efficiency, excluding replica repetition, is defined as the ratio between the expected number of information bits carried by a command and the expected length of a single replica, i.e.,
\begin{equation}
\bar{\eta}_{\mathrm{info}}(L_{\mathrm{b}})
=
\frac{k\E[I]}{n\E[I]+2L_{\mathrm{b}}}.
\label{eq:average_efficiency}
\end{equation}
Over an observation interval of $T$ symbol times, let $\mathcal{U}(T)$ denote the set of distinct command units transmitted and let $\mathcal{D}(T)\subseteq\mathcal{U}(T)$ contain those correctly recovered after all \ac{SIC} rounds. Each command unit is counted only once in both sets, irrespective of the number of replicas. Since each command $u$ is transmitted through $N_{\mathrm{rep}}$ replicas of length $L_{\iota_u}=\iota_u n+2L_{\mathrm{b}}$, we define the effective payload efficiency as
\begin{equation}
\eta_{\mathrm{eff}}(\lambda,L_{\mathrm{b}})
=
\frac{\E\left[\sum_{u\in\mathcal{D}(T)}\iota_u k\right]}
{\E\left[N_{\mathrm{rep}}\sum_{u\in\mathcal{U}(T)}\left(\iota_u n+2L_{\mathrm{b}}\right)\right]}.
\label{eq:effective_efficiency}
\end{equation}
The denominator counts all transmitted symbols across replicas, including overlaps separately; hence, $\eta_{\mathrm{eff}}$ is an efficiency per transmitted symbol, rather than a throughput per unit time, and accounts for boundary overhead, replica repetition, and unsuccessful recovery. When all command units are correctly recovered, it reduces to $\bar{\eta}_{\mathrm{info}}(L_{\mathrm{b}})/N_{\mathrm{rep}}$.

Another metric we consider is \ac{PLR}, defined as the probability that a transmitted command unit is not accepted as valid after processing all its replicas and completing all \ac{SIC} rounds. Commands accepted as valid but recovered erroneously are instead counted in the \ac{UPER} (see Section~\ref{sec:numres}).

\section{Analytical Impact of Boundary Shortening}
\label{sec:shorten}

Let us study the effect of shortening the boundaries. 

\subsection{Payload Efficiency and Collision Exposure}

Let the number of codewords per command have probability mass function $p_{\iota}=\Pr(I=\iota)$. Relative to the full-length design $L_{\mathrm{b}}=n$, the efficiency gain for a command containing $\iota$ codewords is
\begin{equation}
\Gamma_{\eta}(\iota,L_{\mathrm{b}})=\frac{\iota n+2n}{\iota n+2L_{\mathrm{b}}}.
\label{eq:efficiency_gain}
\end{equation}
For example, for $n=128$, $k=64$, and $\iota=1$, reducing each boundary from $128$ to $32$ symbols doubles the payload efficiency from $1/6$ to $1/3$.

Boundary shortening also reduces the temporal exposure of each replica to interference. To isolate this effect, we consider a discrete-time independent-replica Poisson-rain surrogate in which replica starts occur with mean rate \(\nu=\lambda N_{\mathrm{rep}}\) per symbol time. Replica durations are independent marks distributed as $L=nI+2L_{\mathrm{b}}$, with mean $\bar{L}=n\E[I]+2L_{\mathrm{b}}$.

Consider a tagged replica conditioned on $I=\iota$, with duration $L_{\iota}=\iota n+2L_{\mathrm{b}}$. An interfering replica of duration $\ell$ overlaps it if its starting time falls within $L_{\iota}+\ell-1$ symbol positions. Hence, the number of overlapping replicas is Poisson distributed with mean
\(\mu_{\mathrm{ov}}(\iota,L_{\mathrm{b}})=\nu\left(L_{\iota}+\bar{L}-1\right),\label{eq:mean_overlapping_replicas}\)
and the collision-free probability is
\begin{equation}p_{\mathrm{cf}}(\iota,L_{\mathrm{b}})=\exp\left[-\nu\left(L_{\iota}+\bar{L}-1\right)\right].\label{eq:collision_free}\end{equation}
Relative to full-length boundaries,
\begin{equation}\frac{p_{\mathrm{cf}}(\iota,L_{\mathrm{b}})}{p_{\mathrm{cf}}(\iota,n)}=\exp\left[4\nu(n-L_{\mathrm{b}})\right],\label{eq:collision_gain}\end{equation}
which is independent of $\iota$.

The collision events of replicas belonging to the same command are generally dependent. Under the additional approximation that they are independent, the probability that at least one of the $N_{\mathrm{rep}}$ replicas is collision-free is
\begin{equation}\widetilde{p}_{\mathrm{rec}}^{(\mathrm{cf})}(\iota,L_{\mathrm{b}})=1-\left[1-p_{\mathrm{cf}}(\iota,L_{\mathrm{b}})\right]^{N_{\mathrm{rep}}}.\label{eq:at_least_one}\end{equation}
This quantity characterizes collision exposure rather than the end-to-end \ac{PLR}, since \ac{SIC} may recover commands whose replicas are initially overlapped. Averaging over the command-length distribution, the probability that every replica overlaps at least one interfering transmission is
\begin{equation}\widetilde{p}_{\mathrm{loss}}^{(\mathrm{cf})}(L_{\mathrm{b}})=1-\sum_{\iota=1}^{\iota_{\max}}p_{\iota}\widetilde{p}_{\mathrm{rec}}^{(\mathrm{cf})}(\iota,L_{\mathrm{b}}).\label{eq:collision_only_loss}\end{equation}

\subsection{Channel-Estimation Penalty and Refinement}

Shortening the start sequence reduces the number of known symbols available for initial channel estimation. To isolate this effect, consider an interference-free replica with $N_{\mathrm{R}}$ receive antennas, a channel vector $\mathbf{h}$ constant over the complete command unit, and \ac{AWGN} variance $\sigma_w^2$ per complex receive sample. Let $\mathbf{R}_{\mathrm{s}}\in\mathbb C^{N_{\mathrm R}\times L_{\mathrm s}}$ collect the samples corresponding to the detected start sequence.  The least-squares estimate is
\begin{equation}\widehat{\mathbf{h}}^{(0)}=\frac{\mathbf{R}_{\mathrm{s}}\mathbf{x}_{\mathrm{s}}^{\mathsf H}}{L_{\mathrm{s}}},\label{eq:initial_estimate}\end{equation}
with
\begin{equation}\E\left[\left\|\widehat{\mathbf{h}}^{(0)}-\mathbf{h}\right\|^2\right]=\frac{N_{\mathrm{R}}\sigma_w^2}{L_{\mathrm{s}}}.\label{eq:initial_mse}\end{equation}
Hence, relative to the full-length design $L_{\mathrm{s}}=n$, the initial \ac{MSE} increases by a factor $n/L_{\mathrm{s}}$.

In the simulated receiver, all codewords of a command are decoded using the initial channel estimate. Channel refinement is performed only after decoding, to reconstruct the replicas for \ac{SIC}. To quantify its potential benefit, we consider an idealized estimator that combines the start sequence with successfully decoded codewords. Admitting only codewords that satisfy all parity checks, the estimate after $q$ decoded codewords is
\begin{equation}\widehat{\mathbf{h}}^{(q)}=\frac{\mathbf{R}_{\mathrm{s}}\mathbf{x}_{\mathrm{s}}^{\mathsf H}+\sum_{j=1}^{q}\mathbf{R}_{j}\widehat{\mathbf{c}}_{j}^{\mathsf H}}{L_{\mathrm{s}}+qn},\label{eq:refinement}\end{equation}
where $\mathbf R_j\in\mathbb C^{N_{\mathrm R}\times n}$ contains the received samples for the $j$th codeword, and $\widehat{\mathbf c}_j$ is its reconstructed modulated sequence. To isolate the estimation gain, assume that the first $q$ codewords are always reconstructed correctly, so that $\widehat{\mathbf c}_j=\mathbf c_j$ for $1\leq j\leq q$. For fixed $q$, the resulting idealized estimator satisfies
\begin{equation}
\E\left[\left\|\widehat{\mathbf h}^{(q)}-\mathbf h\right\|^2\right]=\frac{N_{\mathrm R}\sigma_w^2}{L_{\mathrm s}+qn}.
\label{eq:refined_mse}
\end{equation}
Under this ideal model, the \ac{MSE} therefore scales as $(L_{\mathrm s}+qn)^{-1}$, showing how correctly reconstructed payload symbols reduce the initial estimation penalty. This idealized expression does not characterize the probability that refinement can be initiated: shortening $L_{\mathrm s}$ may itself prevent successful decoding of the first codeword. Moreover, when the candidate window overlaps an interferer, the initial estimate contains an additional interference projection onto the start sequence.

\section{Deep-Learning-Based Length-Decoupled Receiver}
\label{sec:receiver}
\begin{table}[t]
  \caption{CNN Architecture and Complexity ($n=128$, $N_{\mathrm{it}}=20$). }
  \label{tab:CNN_compl}
  \centering
  \scriptsize
    \begin{tabular}{l|l|r|r|r|r}
    \toprule
    \textbf{Layer type} & \textbf{Layer size} & \multicolumn{4}{c}{\textbf{Complexity} [kFLOPs] ($L_{\text{b}}$)} \\
     & & $16$ & $32$ & $64$ & $128$ \\
    \midrule
    \multicolumn{6}{l}{\textit{Branch 1}} \\
    Input $\bm Y_1$   & $3 \times 2 L_{\text{b}}$ & --   & --   & --   & --   \\
    2D convolutional  & $8$@$2\times 7$          & $11.6$  & $26.0$  & $54.7$  & $112.0$  \\
    Fully connected   & $130$                    & $108.3$ & $241.4$ & $507.7$ & $1040.1$ \\
    Fully connected   & $65$                     & $17.0$  & $17.0$  & $17.0$  & $17.0$   \\
    \midrule
    \multicolumn{6}{l}{\textit{Branch 2}} \\
    Input $\bm Y_2$   & $(N_{\mathrm{it}}{+}1) n$   & --   & --   & --   & --   \\
    Fully connected   & $130$                    & $699.0$ & $699.0$ & $699.0$ & $699.0$ \\
    Fully connected   & $65$                     & $17.0$  & $17.0$  & $17.0$  & $17.0$  \\
    \midrule
    \multicolumn{6}{l}{\textit{Merged Backbone}} \\
    Fully connected   & $130$                    & $33.9$ & $33.9$ & $33.9$ & $33.9$ \\
    Fully connected   & $65$                     & $17.0$ & $17.0$ & $17.0$ & $17.0$ \\
    Fully connected   & $2$                      & $0.3$  & $0.3$  & $0.3$  & $0.3$  \\
    Output            & $2$                      & $<0.1$ & $<0.1$ & $<0.1$ & $<0.1$ \\
    \bottomrule
    \end{tabular}
\end{table}
We propose a unified \ac{DL} framework for the joint detection of start and tail sequences, using signal-level features for start detection and decoder information for tail detection. Unlike the  design of~\cite{BattaglioniCSCN2025,BattaglioniArxiv2026}, the proposed \ac{CNN} is \emph{length-decoupled}: the boundary length  is a design parameter independent of the codeword length $n$ processed by the \ac{LDPC} decoder. This allows $L_{\mathrm{b}}<n$, reducing framing overhead and collision exposure (Section~\ref{sec:shorten}), while the decoder keeps operating on $n$-symbol blocks irrespective of $L_{\mathrm{b}}$.

The architecture (Table~\ref{tab:CNN_compl}) is multi-branch and multi-label: two parallel branches process the inputs $\bm{Y}_1$ and $\bm{Y}_2$ and merge into a shared fully connected stage, whose sigmoid outputs independently represent the probabilities of the start ($A$) and tail ($B$), with labels $[A,B]\in\{0,1\}^2$ not mutually exclusive. The training samples are organized into five input categories: $H_0:[A,B]=[0,0]$, noise or non-informative fragments; $H_1:[1,0]$, start sequence; $H_2:[0,1]$, tail sequence; $H_3:[0,0]$, codeword;\footnote{Although $H_0$ and $H_3$ share the label $[0,0]$, they are retained as distinct training categories so that the network is exposed to both non-informative fragments and valid codewords as negative examples.} $H_4:[1,1]$, overlapping start and tail.

\subsubsection{Branch 1} It operates on windows of length $L_{\mathrm{b}}$, matched to the start/tail sequences. Its input $\bm{Y}_1 \in \mathbb{R}^{3 \times (2 L_{\mathrm{b}})}$ concatenates the real and imaginary parts of the known start sequence, of the row vector of \ac{MRC} samples $\hat{\mathbf{x}}\in\mathbb{C}^{1\times L_{\mathrm b}}$, and of the known tail sequence:
\begin{align}
    \bm{Y}_1 & =
    \setlength\arraycolsep{2pt}
    \begin{bmatrix}
\Re(\mathbf{x}_{\mathrm{s}}) & \Im(\mathbf{x}_{\mathrm{s}}) \\
\Re(\hat{\mathbf{x}}) & \Im(\hat{\mathbf{x}}) \\
\Re(\mathbf{x}_{\mathrm{t}}) & \Im(\mathbf{x}_{\mathrm{t}})
\end{bmatrix}
    \,.
    \label{eq:Y1}
\end{align}
Although the known \ac{BPSK} sequences satisfy $\Im(\mathbf x_{\mathrm s})=\Im(\mathbf x_{\mathrm t})=\mathbf 0$, these entries are retained to provide a common real-imaginary layout for the complex \ac{MRC} estimate $\widehat{\mathbf x}$, whose imaginary component reflects noise, interference, and channel-estimation errors. During start detection, the \ac{MRC} observation is formed using the provisional channel estimate obtained from the current candidate window under the start-sequence hypothesis. Since $L_{\mathrm{b}}$ is decoupled from $n$, shrinking it directly reduces the size of $\bm{Y}_1$ and the cost of Branch~1, as quantified in Section~\ref{sec:receiver}.

\subsubsection{Branch 2} It has a fixed width $n$, matched to the decoder block length. Its input is $\bm{Y}_2 \in \mathbb{R}^{(N_{\mathrm{it}}+1) \times n}$,
\begin{align}
    \bm{Y}_2 & =
    \setlength\arraycolsep{2pt}
    \begin{bmatrix}
        \bm{L} \\
        \bm{f}
    \end{bmatrix},
    \label{eq:Y2}
\end{align}
where $\bm{Y}_2$ has different contents for start and tail detection. For start detection, $\bm{L}\in\mathbb{R}^{N_{\mathrm{it}}\times n}$ is filled by replicating, over all entries, the average normalized correlation (across the $N_{\mathrm{R}}$ receive antennas) between the current $L_{\mathrm{b}}$-symbol observation window and the known start sequence, while $\bm{f}=\mathbf{0}_{n}$. For tail detection, $\bm{L}$ collects the \acp{LLR} produced by the decoder over its $N_{\mathrm{it}}$ iterations, and $\bm{f} \in \{ \mathbf{0}_{n}, \mathbf{1}_{n} \}$ is a flag vector indicating whether the decoder converged.

For each $L_{\mathrm{b}}$, a single trained network performs both start and tail detection. During start detection, the receiver has no synchronization reference: a window of length $L_{\mathrm{b}}$ slides with unit step; after synchronization, tail detection slides a window of length $n$ with step $n$, aligned to successive decoder blocks, and Branch~2 also exploits decoder information. In both phases, $\bm{Y}_2$ retains width $n$, while $\bm{Y}_1$ contains the first $L_{\mathrm b}$ samples of the observation window. For $Z\in\{A,B\}$, $R_Z=\Pr(\widehat Z=1\mid Z=1)$ and $F_Z=\Pr(\widehat Z=1\mid Z=0)$ denote per-window detection and false-alarm probabilities. A start declaration only creates a candidate, which enters \ac{SIC} after successful decoding and boundary validation. Consequently, false starts are accounted for in the end-to-end simulation and primarily increase the number of attempted decoding operations rather than directly causing packet loss.

We evaluate the \ac{CNN} inference cost per tested window in \acp{FLOP}. A convolutional layer requires $C_{\mathrm{cv}} = 2\, N_{\mathrm{cv}}\cdot F_{\mathrm{cv}}\cdot G_{\mathrm{cv}}\cdot D_{\mathrm{cv}}$ FLOPs, where $N_{\mathrm{cv}}$, $F_{\mathrm{cv}}$, $G_{\mathrm{cv}}$, and $D_{\mathrm{cv}}$ are the number of filters, the filter size, the number of channels, and the output size, respectively; the \ac{ReLU} adds $C_{\mathrm{ReLU}} = N_{\mathrm{cv}}\cdot D_{\mathrm{cv}}$, a fully connected layer requires $C_{\mathrm{FC}} = 2\,\mathrm{in}\cdot\mathrm{out} + \mathrm{out}$, and the final sigmoid with thresholding costs $4$ FLOPs per label. Per-layer counts are reported in Table~\ref{tab:CNN_compl}. For $n=128$ and $N_{\mathrm{it}}=20$, the \ac{CNN} forward-pass cost is approximately $0.91$, $1.05$, $1.35$, and $1.94$~MFLOP for $L_{\mathrm b}=16$, $32$, $64$, and $128$, respectively. 

\section{Numerical Results}
\label{sec:numres}

We adopt the indoor industrial setting of~\cite{BattaglioniArxiv2026}: one transmit antenna and $N_{\mathrm{R}}=4$ receive antennas, two replicas per virtual frame, and a $(128,64)$ \ac{LDPC} code decoded by \ac{NMS} with $N_{\mathrm{it}}=20$; the number of codewords per command is uniform over $[1,\iota_{\max}]$, $\iota_{\max}=5$. All propagation, transmit-power, receiver-noise, and \ac{SIC} parameters not explicitly reported here for space reasons are identical to those in~\cite{BattaglioniArxiv2026}. Each virtual frame contains $N_{\mathrm S}=10$ slots, with replica placement as described in Section~\ref{sec:sysmod}.  We consider $L_{\mathrm{b}}\in\{16,32,64,128\}$, with $L_{\mathrm{b}}=128$ the full-length case of~\cite{BattaglioniCSCN2025,BattaglioniArxiv2026}. Under the above slotting rule, shortening $L_{\mathrm b}$ also reduces the virtual-frame duration from $8960$ to $6720$ symbols and compresses the replica-spreading interval; this temporal effect is included in the end-to-end results but not in the Poisson-rain surrogate.

To obtain the following results, for each $L_{\mathrm{b}}\in\{16,32,64,128\}$, a dedicated \ac{CNN} is trained using data generated with the corresponding boundary configuration and a common activation rate of $\lambda=10^{-3}$, selected to reproduce the reference indoor industrial operating scenario considered in this work. For each configuration, up to $50\,000$ samples are collected for each of the classes $H_0$--$H_3$, whereas the $H_4$ class, corresponding to coincident start- and tail-sequence events, is naturally less frequent under the considered traffic conditions and therefore contains fewer samples. The resulting dataset is split $70/30$ between training and validation sets. Training uses the binary cross-entropy loss and the Adam optimizer, with dropout $0.2$, a mini-batch size of $50$, and a learning rate of $10^{-3}$.  All reported results are obtained from independent realizations not used for training or validation.

\begin{figure}[t]
    \centering
    \resizebox{0.85\columnwidth}{!}{%
        \definecolor{mycolor1}{rgb}{0.929,0.694,0.125} 
\definecolor{mycolor2}{rgb}{0.494,0.184,0.556} 
\definecolor{mycolor3}{rgb}{0.000,0.447,0.741} 
\definecolor{mycolor4}{rgb}{0.850,0.325,0.098} 

\begin{tikzpicture}
\begin{axis}[
    width=\columnwidth,
    height=0.7\columnwidth,
    scale only axis,
    xmin=1e-4,
    xmax=2e-3,
    scaled x ticks=false,
    ymin=1e-2,
    ymax=1,
    ymode=log,
    xlabel={$\lambda\,[10^{-3}]$},
    ylabel={$\tilde{p}_{\mathrm{loss}}^{(\mathrm{cf})}$},
    xtick={1e-4,5e-4,1e-3,1.5e-3,2e-3},
    xticklabels={$0.1$,$0.5$,$1$,$1.5$,$2$},
    ytick={0.01,0.1,1},
    tick label style={font=\normalsize},
label style={font=\large},
    major grid style={draw=gray!25},
    xmajorgrids,
    ymajorgrids,
    axis line style={black},
    tick style={black},
    legend style={
        at={(0.47,0.35)},
        anchor=north west,
        draw=gray!40,
        rounded corners=1pt,
        fill=white,
        fill opacity=0.88,
        text opacity=1,
        font=\normalsize,
        row sep=-1pt
    },
]

\addplot[
    color=mycolor1,
    line width=1.1pt,
    mark=none
]
table[row sep=crcr]{
0.0001 0.024217971677427\\
0.0002 0.0816223058375244\\
0.0003 0.155735276707866\\
0.0004 0.236240947431871\\
0.0005 0.316860497490065\\
0.0006 0.393940417943306\\
0.0007 0.465516192949251\\
0.0008 0.530693173390228\\
0.0009 0.589238987319026\\
0.0010 0.641316910459722\\
0.0011 0.687313024408355\\
0.0012 0.727725623447033\\
0.0013 0.763095784293912\\
0.0014 0.793965009344791\\
0.0015 0.820850539814211\\
0.0016 0.844232075132445\\
0.0017 0.864545739325971\\
0.0018 0.882182545045335\\
0.0019 0.897489549764371\\
0.0020 0.910772529523359\\
};
\addlegendentry{$L_{\mathrm b}=16$}

\addplot[
    color=mycolor1,
    only marks,
    mark=o,
    mark size=1.4pt,
    line width=0.7pt,
    forget plot
]
table[row sep=crcr]{
0.0001 0.0387242138994592\\
0.0002 0.10085151942685\\
0.0003 0.173658692951877\\
0.0004 0.254918426103647\\
0.0005 0.330411166803123\\
0.0006 0.399023457162992\\
0.0007 0.469206215791633\\
0.0008 0.532864249181772\\
0.0009 0.584785926580448\\
0.0010 0.636841525567983\\
0.0011 0.6840007211394\\
0.0012 0.719867715588515\\
0.0013 0.760122268991782\\
0.0014 0.78745148381791\\
0.0015 0.817259086095626\\
0.0016 0.840431793920166\\
0.0017 0.858532626664396\\
0.0018 0.876596936682774\\
0.0019 0.89178509462536\\
0.0020 0.905696972133174\\
};

\addplot[
    color=mycolor2,
    line width=1.1pt,
    mark=none
]
table[row sep=crcr]{
0.0001 0.0275991116864156\\
0.0002 0.0920288037085565\\
0.0003 0.1738487390476\\
0.0004 0.261282087828502\\
0.0005 0.347440215619026\\
0.0006 0.428523185628211\\
0.0007 0.502656019639933\\
0.0008 0.56913912787732\\
0.0009 0.627968378392785\\
0.0010 0.679530470183703\\
0.0011 0.724412009585532\\
0.0012 0.76328206394223\\
0.0013 0.796821937282896\\
0.0014 0.825685051764446\\
0.0015 0.850475800213626\\
0.0016 0.871740150714727\\
0.0017 0.889963346509221\\
0.0018 0.905571719514203\\
0.0019 0.918936728692918\\
0.0020 0.930380045425753\\
};
\addlegendentry{$L_{\mathrm b}=32$}

\addplot[
    color=mycolor2,
    only marks,
    mark=o,
    mark size=1.4pt,
    line width=0.7pt,
    forget plot
]
table[row sep=crcr]{
0.0001 0.0430463244213207\\
0.0002 0.111787764244715\\
0.0003 0.193369224137067\\
0.0004 0.27690451018332\\
0.0005 0.360708021049274\\
0.0006 0.435895396241424\\
0.0007 0.502583329982711\\
0.0008 0.569517300640468\\
0.0009 0.626347641377099\\
0.0010 0.677344735735315\\
0.0011 0.719851502934015\\
0.0012 0.762163828771483\\
0.0013 0.791405944955598\\
0.0014 0.821592832254853\\
0.0015 0.847509379739762\\
0.0016 0.867602958816473\\
0.0017 0.884070796460177\\
0.0018 0.90230732446003\\
0.0019 0.91667165101581\\
0.0020 0.927171454478422\\
};

\addplot[
    color=mycolor3,
    line width=1.1pt,
    mark=none
]
table[row sep=crcr]{
0.0001 0.03490890690422\\
0.0002 0.113928749940265\\
0.0003 0.21097778372712\\
0.0004 0.311308828981016\\
0.0005 0.407015131599488\\
0.0006 0.494258941626438\\
0.0007 0.571570135191525\\
0.0008 0.63880894623091\\
0.0009 0.696540142374031\\
0.0010 0.745661320166862\\
0.0011 0.787187378451872\\
0.0012 0.822130255684822\\
0.0013 0.851436121322959\\
0.0014 0.875956632627436\\
0.0015 0.896439866883587\\
0.0016 0.913532147772164\\
0.0017 0.927785472617209\\
0.0018 0.939667408198379\\
0.0019 0.949571653759844\\
0.0020 0.95782828232497\\
};
\addlegendentry{$L_{\mathrm b}=64$}

\addplot[
    color=mycolor3,
    only marks,
    mark=o,
    mark size=1.4pt,
    line width=0.7pt,
    forget plot
]
table[row sep=crcr]{
0.0001 0.0513595166163142\\
0.0002 0.1337374749499\\
0.0003 0.229737830708976\\
0.0004 0.327481471876153\\
0.0005 0.4133167342677\\
0.0006 0.493193193193193\\
0.0007 0.573952027356137\\
0.0008 0.630038358805455\\
0.0009 0.694026798096397\\
0.0010 0.740562273035636\\
0.0011 0.781328805521787\\
0.0012 0.818505444836655\\
0.0013 0.846323698382407\\
0.0014 0.871227686849148\\
0.0015 0.888269451781358\\
0.0016 0.907464478687212\\
0.0017 0.922911732830249\\
0.0018 0.936259007205765\\
0.0019 0.946435565017423\\
0.0020 0.955562001160209\\
};

\addplot[
    color=mycolor4,
    line width=1.1pt,
    mark=none
]
table[row sep=crcr]{
0.0001 0.0514996681872821\\
0.0002 0.161006388440447\\
0.0003 0.286694890228241\\
0.0004 0.408236684881697\\
0.0005 0.516837723550493\\
0.0006 0.609706673853871\\
0.0007 0.687016346554596\\
0.0008 0.750263088892075\\
0.0009 0.801405280286141\\
0.0010 0.842431745562416\\
0.0011 0.875163684610461\\
0.0012 0.90118026412428\\
0.0013 0.921806982498058\\
0.0014 0.938133544554493\\
0.0015 0.951043490204259\\
0.0016 0.96124645379198\\
0.0017 0.969308673019472\\
0.0018 0.975679916827883\\
0.0019 0.980716326886605\\
0.0020 0.984699315955322\\
};
\addlegendentry{$L_{\mathrm b}=128$}

\addplot[
    color=mycolor4,
    only marks,
    mark=o,
    mark size=1.4pt,
    line width=0.7pt,
    forget plot
]
table[row sep=crcr]{
0.0001 0.0705599984065491\\
0.0002 0.18236478970628\\
0.0003 0.300568545803972\\
0.0004 0.41822595630326\\
0.0005 0.519381626677592\\
0.0006 0.607086920521543\\
0.0007 0.678398009550945\\
0.0008 0.745228530887143\\
0.0009 0.7955635671724\\
0.0010 0.835718494402445\\
0.0011 0.868576807907624\\
0.0012 0.893605076640836\\
0.0013 0.919153732593863\\
0.0014 0.932556136450133\\
0.0015 0.948297352545246\\
0.0016 0.957762442576581\\
0.0017 0.966509565842553\\
0.0018 0.974842829858348\\
0.0019 0.978986464840621\\
0.0020 0.982747040129924\\
};

\end{axis}
\end{tikzpicture}
    }
    \caption{Probability that every replica overlaps at least one interfering transmission versus activation rate for different $L_{\mathrm{b}}$. Solid lines denote the Poisson-rain approximation in \eqref{eq:collision_only_loss}, while markers show Monte Carlo results.}
    \label{fig:collision}
\end{figure}
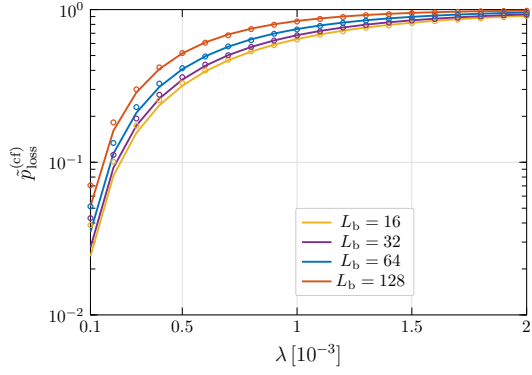

For the considered command-length distribution, $\E[I]=3$, and \eqref{eq:average_efficiency} gives $\bar{\eta}_{\mathrm{info}}=0.462$, $0.429$, $0.375$, and $0.300$ for $L_{\mathrm{b}}=16$, $32$, $64$, and $128$, respectively. Thus, the shortest boundary increases the average payload efficiency by $53.8\%$ relative to the full-length design.  Fig.~\ref{fig:collision} compares \eqref{eq:collision_only_loss} with Monte Carlo results preserving the virtual-frame placement rule. The surrogate captures the relative ordering and traffic trend, confirming that shorter boundaries reduce the probability that all replicas overlap.

As a conventional benchmark, we use the \ac{GLRT} detector of~\cite{Khan2024} for start detection. Treating the unknown per-antenna channel gains as nuisance parameters and assuming known noise variance yields the noncoherent statistic
\[
\Lambda(\mathbf R)=\sum_{m=1}^{N_{\mathrm R}}\left|\sum_{i=1}^{L_{\mathrm b}}r_{m,i}x_{\mathrm{s},i}^{*}\right|^2,
\]
where $\mathbf R=[r_{m,i}]\in\mathbb C^{N_{\mathrm R}\times L_{\mathrm b}}$ contains the received samples in the candidate window. For tail detection, the benchmark instead uses the channel estimate obtained from the start sequence to form the \ac{MRC} output and correlates its first $L_{\mathrm b}$ samples with the known tail sequence, using
\[
\Lambda_{\mathrm t}=\left|\sum_{i=1}^{L_{\mathrm b}}\Re\{\widehat{x}_i\}x_{\mathrm t,i}\right|^2.
\]

We refer to this combination of \ac{GLRT}-based start detection and correlation-based tail detection as the \ac{GLRT} benchmark. Both receivers use the same truncated boundary sequences. This provides a comparison under a common sequence construction, although sequences optimized separately for each $L_{\mathrm b}$ could improve conventional detection at short boundary lengths~\cite{CoxsonRusso2005}. As shown in Fig.~\ref{fig:ROC}, the proposed receiver achieves higher detection probabilities than the benchmark. For start detection, performance degrades as $L_{\mathrm b}$ decreases because fewer known symbols are available. For tail detection, the proposed \ac{CNN} additionally exploits the decoder \ac{LLR} trajectory and convergence flag, and therefore exhibits a much weaker dependence on $L_{\mathrm b}$.

\begin{figure}[!t]
  \centering

  \ifcsname subfloat\endcsname
    \def\rocsub#1#2#3{\subfloat[#1\label{#2}]{#3}}%
  \else\ifcsname subfigure\endcsname
    \def\rocsub#1#2#3{\begin{subfigure}{\columnwidth}\centering #3%
      \caption{#1}\label{#2}\end{subfigure}}%
  \else
    \ifcsname c@rocsubfig\endcsname\else\newcounter{rocsubfig}\fi
    \setcounter{rocsubfig}{0}\renewcommand\therocsubfig{\alph{rocsubfig}}%
    \expandafter\def\csname p@rocsubfig\endcsname{\the\numexpr\value{figure}+1\relax}%
    \def\rocsub#1#2#3{\par\refstepcounter{rocsubfig}\label{#2}%
      {\centering #3\par\vspace{1mm}\footnotesize(\therocsubfig) #1\par}}%
  \fi\fi
  
  \definecolor{rocred}{RGB}{237,33,41}
  \definecolor{rocblue}{RGB}{30,144,255}
  \definecolor{rocpurple}{RGB}{128,0,128}
  \definecolor{rocgold}{RGB}{255,215,0}

  \pgfplotsset{
    rocdash/.style={dash pattern=on 3.5pt off 1.5pt},
    rocstyle/.style={
      scale only axis,
      width=0.80\columnwidth,
      height=0.25\columnwidth,
      xmin=0, xmax=0.30,
      ymin=0.80, ymax=1.00,
      xtick={0,0.05,...,0.30},
      ytick={0.80,0.85,...,1.00},
      xticklabel style={/pgf/number format/.cd, fixed, fixed zerofill, precision=2},
      yticklabel style={/pgf/number format/.cd, fixed, fixed zerofill, precision=2},
      tick label style={font=\scriptsize},
      label style={font=\small},
      xlabel={False Alarm Rate ($F$)},
      ylabel={Detection Rate ($R$)},
      xlabel style={yshift=3pt},
      ylabel style={yshift=-3pt},
      grid=major,
      grid style={gray!35, line width=0.3pt},
      tick align=outside,
      tick pos=left,
      clip=true,
      every axis plot/.append style={line width=0.9pt, no markers},
      legend columns=4, 
      transpose legend,
      legend pos=south east,
      legend cell align=left,
      legend style={font=\tiny, draw=gray!40, rounded corners=1pt,
                    row sep=-1.5pt, column sep=4pt,
                    /tikz/every even column/.append style={column sep=8pt}},
      legend image post style={line width=0.9pt},
    },
  }

  \rocsub{Start sequence performance.}{fig:roc_start}{%
    \begin{tikzpicture}
      \begin{axis}[rocstyle]
        \addplot[rocred] coordinates {(0.00005,0.73770) (0.00079,0.90990) (0.00158,0.92941) (0.00237,0.93889) (0.00316,0.94383) (0.00395,0.94737) (0.00553,0.95257) (0.00711,0.95717) (0.00868,0.96113) (0.01026,0.96440) (0.01105,0.96577) (0.01263,0.96799) (0.01500,0.97087) (0.01737,0.97339) (0.01974,0.97556) (0.02211,0.97741) (0.02447,0.97892) (0.02684,0.98013) (0.03158,0.98205) (0.03632,0.98370) (0.04184,0.98533) (0.04737,0.98668) (0.05368,0.98795) (0.06158,0.98924) (0.07737,0.99144) (0.09158,0.99314) (0.10500,0.99449) (0.11842,0.99556) (0.13105,0.99628) (0.14447,0.99675) (0.18789,0.99793) (0.22895,0.99876) (0.26763,0.99927) (0.30474,0.99949) (0.31193,0.99950)};
        \addlegendentry{CNN, $L_\mathrm{b}=128$}
        \addplot[rocred, rocdash] coordinates {(0.01010,0.73770) (0.01184,0.77458) (0.01263,0.78874) (0.01342,0.80050) (0.01500,0.82099) (0.01658,0.83841) (0.01816,0.85375) (0.01974,0.86808) (0.02132,0.88117) (0.02211,0.88708) (0.02289,0.89249) (0.02368,0.89733) (0.02447,0.90153) (0.02605,0.90887) (0.02763,0.91528) (0.02921,0.92096) (0.03079,0.92608) (0.03316,0.93311) (0.03632,0.94204) (0.03868,0.94835) (0.04026,0.95225) (0.04184,0.95584) (0.04342,0.95906) (0.04500,0.96183) (0.04658,0.96410) (0.04895,0.96708) (0.05132,0.96974) (0.05368,0.97212) (0.05605,0.97424) (0.05921,0.97673) (0.06237,0.97891) (0.06711,0.98180) (0.07184,0.98441) (0.07658,0.98671) (0.08053,0.98835) (0.08447,0.98972) (0.08842,0.99081) (0.09553,0.99244) (0.10263,0.99382) (0.10974,0.99494) (0.11684,0.99580) (0.12553,0.99655) (0.13974,0.99749) (0.15553,0.99825) (0.17447,0.99884) (0.20526,0.99955) (0.22816,0.99981) (0.30553,1.00000) (0.31193,1.00000)};
        \addlegendentry{GLRT, $L_\mathrm{b}=128$}
        \addplot[rocblue] coordinates {(0.00037,0.73770) (0.00079,0.89741) (0.00158,0.90838) (0.00237,0.91361) (0.00474,0.93490) (0.00553,0.93951) (0.00711,0.94671) (0.00789,0.94983) (0.00868,0.95262) (0.00947,0.95508) (0.01026,0.95719) (0.01105,0.95894) (0.01184,0.96033) (0.01342,0.96251) (0.01500,0.96442) (0.01658,0.96609) (0.01895,0.96822) (0.02132,0.96996) (0.02368,0.97143) (0.02763,0.97354) (0.03316,0.97621) (0.03868,0.97859) (0.04421,0.98067) (0.04895,0.98218) (0.05368,0.98344) (0.05842,0.98442) (0.06395,0.98524) (0.07500,0.98659) (0.08763,0.98783) (0.10184,0.98894) (0.12000,0.99005) (0.16737,0.99252) (0.21158,0.99463) (0.24474,0.99594) (0.27395,0.99681) (0.30000,0.99732) (0.31193,0.99745)};
        \addlegendentry{CNN, $L_\mathrm{b}=64$}
        \addplot[rocblue, rocdash] coordinates {(0.01604,0.73770) (0.01737,0.75664) (0.01974,0.78688) (0.02132,0.80534) (0.02289,0.82175) (0.02368,0.82899) (0.02526,0.84151) (0.02684,0.85265) (0.02842,0.86272) (0.03000,0.87183) (0.03158,0.88013) (0.03316,0.88776) (0.03553,0.89838) (0.03789,0.90830) (0.03947,0.91445) (0.04105,0.92017) (0.04263,0.92540) (0.04421,0.93010) (0.04579,0.93422) (0.04816,0.93956) (0.05053,0.94444) (0.05289,0.94890) (0.05526,0.95293) (0.05763,0.95655) (0.06000,0.95977) (0.06237,0.96260) (0.06474,0.96507) (0.06789,0.96810) (0.07105,0.97087) (0.07421,0.97338) (0.07737,0.97565) (0.08053,0.97768) (0.08447,0.97989) (0.08842,0.98176) (0.09395,0.98406) (0.09947,0.98612) (0.10579,0.98816) (0.11211,0.98989) (0.11842,0.99133) (0.12711,0.99294) (0.13816,0.99475) (0.14763,0.99605) (0.15632,0.99698) (0.16500,0.99763) (0.18316,0.99850) (0.20289,0.99921) (0.22184,0.99961) (0.24711,0.99979) (0.30632,1.00000) (0.31193,1.00000)};
        \addlegendentry{GLRT, $L_\mathrm{b}=64$}
        \addplot[rocpurple] coordinates {(0.00067,0.73770) (0.00079,0.79024) (0.00158,0.80554) (0.00237,0.83799) (0.00316,0.85384) (0.00395,0.86659) (0.00474,0.87614) (0.00553,0.88251) (0.00632,0.88744) (0.00711,0.89164) (0.00789,0.89524) (0.00868,0.89837) (0.01026,0.90375) (0.01263,0.91155) (0.01658,0.92485) (0.01895,0.93230) (0.02053,0.93685) (0.02211,0.94097) (0.02368,0.94455) (0.02526,0.94752) (0.02605,0.94875) (0.02763,0.95066) (0.03079,0.95393) (0.03395,0.95686) (0.03711,0.95947) (0.04026,0.96180) (0.04342,0.96385) (0.04737,0.96609) (0.05132,0.96799) (0.05526,0.96961) (0.06000,0.97124) (0.06711,0.97332) (0.07579,0.97556) (0.08447,0.97754) (0.09395,0.97941) (0.10421,0.98115) (0.11605,0.98286) (0.12947,0.98451) (0.15079,0.98679) (0.18789,0.99054) (0.21395,0.99290) (0.23605,0.99463) (0.25579,0.99590) (0.27395,0.99681) (0.29132,0.99740) (0.30789,0.99768) (0.31193,0.99770)};
        \addlegendentry{CNN, $L_\mathrm{b}=32$}
        \addplot[rocpurple, rocdash] coordinates {(0.03187,0.73770) (0.03395,0.75165) (0.03632,0.76674) (0.03868,0.78092) (0.04105,0.79432) (0.04342,0.80705) (0.04658,0.82319) (0.05289,0.85474) (0.05526,0.86593) (0.05763,0.87638) (0.05921,0.88282) (0.06079,0.88874) (0.06237,0.89407) (0.06395,0.89876) (0.06632,0.90525) (0.06868,0.91135) (0.07105,0.91707) (0.07342,0.92242) (0.07579,0.92739) (0.07816,0.93200) (0.08053,0.93623) (0.08289,0.94011) (0.08526,0.94362) (0.08763,0.94678) (0.09158,0.95157) (0.09553,0.95601) (0.09947,0.96008) (0.10342,0.96380) (0.10737,0.96718) (0.11132,0.97020) (0.11526,0.97288) (0.11921,0.97523) (0.12395,0.97768) (0.13105,0.98109) (0.13737,0.98383) (0.14368,0.98628) (0.14921,0.98816) (0.15474,0.98978) (0.16026,0.99113) (0.16579,0.99218) (0.17842,0.99406) (0.19026,0.99556) (0.20211,0.99678) (0.21316,0.99765) (0.22421,0.99826) (0.23921,0.99874) (0.27000,0.99949) (0.29684,0.99988) (0.31193,0.99996)};
        \addlegendentry{GLRT, $L_\mathrm{b}=32$}
        \addplot[rocgold] coordinates {(0.00285,0.73770) (0.00316,0.74232) (0.00395,0.74950) (0.00474,0.75513) (0.00553,0.76223) (0.00947,0.80192) (0.01105,0.81601) (0.01263,0.82796) (0.01579,0.85053) (0.01816,0.86642) (0.01974,0.87620) (0.02132,0.88512) (0.02289,0.89303) (0.02447,0.89974) (0.02526,0.90260) (0.02605,0.90509) (0.02684,0.90720) (0.02921,0.91245) (0.03158,0.91733) (0.03395,0.92185) (0.03632,0.92603) (0.03868,0.92988) (0.04105,0.93340) (0.04342,0.93662) (0.04579,0.93952) (0.04816,0.94214) (0.05053,0.94446) (0.05289,0.94651) (0.05526,0.94829) (0.05763,0.94982) (0.06000,0.95109) (0.06316,0.95247) (0.06868,0.95462) (0.07500,0.95678) (0.08132,0.95867) (0.08842,0.96050) (0.09632,0.96225) (0.10579,0.96404) (0.11921,0.96627) (0.13737,0.96930) (0.18711,0.97827) (0.21316,0.98262) (0.23684,0.98630) (0.25816,0.98934) (0.27868,0.99198) (0.29763,0.99416) (0.31193,0.99562)};
        \addlegendentry{CNN, $L_\mathrm{b}=16$}
        \addplot[rocgold, rocdash] coordinates {(0.07268,0.73770) (0.07579,0.75161) (0.07816,0.76170) (0.08053,0.77128) (0.08289,0.78033) (0.08526,0.78879) (0.08763,0.79661) (0.09079,0.80633) (0.09395,0.81552) (0.09711,0.82422) (0.10026,0.83247) (0.10342,0.84031) (0.10658,0.84779) (0.11053,0.85668) (0.11447,0.86514) (0.11921,0.87483) (0.13105,0.89831) (0.13658,0.90895) (0.14132,0.91768) (0.14526,0.92457) (0.14842,0.92976) (0.15158,0.93464) (0.15474,0.93915) (0.15789,0.94325) (0.16026,0.94604) (0.16263,0.94855) (0.16500,0.95078) (0.16816,0.95337) (0.17368,0.95764) (0.17921,0.96161) (0.18474,0.96529) (0.19026,0.96868) (0.19579,0.97179) (0.20132,0.97463) (0.20684,0.97719) (0.21237,0.97947) (0.21789,0.98150) (0.22342,0.98325) (0.22895,0.98475) (0.24632,0.98893) (0.25895,0.99170) (0.27000,0.99384) (0.27947,0.99540) (0.28816,0.99657) (0.29684,0.99746) (0.30474,0.99800) (0.31193,0.99823)};
        \addlegendentry{GLRT, $L_\mathrm{b}=16$}
      \end{axis}
    \end{tikzpicture}}

  \vspace{2mm}
  \rocsub{Tail sequence performance.}{fig:roc_tail}{%
    \begin{tikzpicture}
      \begin{axis}[rocstyle]
        \addplot[rocred] coordinates {(0.00059,0.73770) (0.00079,0.97695) (0.00158,0.98118) (0.00237,0.98361) (0.00316,0.98538) (0.00395,0.98663) (0.00474,0.98770) (0.00553,0.98800) (0.00711,0.98835) (0.00947,0.98922) (0.01342,0.99097) (0.01658,0.99180) (0.01974,0.99230) (0.02447,0.99277) (0.03316,0.99328) (0.06079,0.99497) (0.06947,0.99501) (0.31193,0.99503)};
        \addlegendentry{CNN, $L_\mathrm{b}=128$}
        \addplot[rocred, rocdash] coordinates {(0.00203,0.73770) (0.00237,0.92692) (0.00316,0.95725) (0.00395,0.96804) (0.00474,0.97426) (0.00553,0.97887) (0.00632,0.98248) (0.00711,0.98524) (0.00789,0.98766) (0.00868,0.98956) (0.00947,0.99063) (0.01026,0.99122) (0.01184,0.99199) (0.01421,0.99319) (0.01658,0.99436) (0.01816,0.99481) (0.02132,0.99529) (0.02605,0.99598) (0.03237,0.99710) (0.03632,0.99737) (0.04500,0.99768) (0.06079,0.99788) (0.08132,0.99826) (0.10974,0.99868) (0.13737,0.99885) (0.31193,0.99942)};
        \addlegendentry{GLRT, $L_\mathrm{b}=128$}
        \addplot[rocblue] coordinates {(0.00061,0.73770) (0.00079,0.96108) (0.00158,0.97398) (0.00237,0.97804) (0.00316,0.98005) (0.00395,0.98179) (0.00474,0.98310) (0.00553,0.98388) (0.01026,0.98712) (0.01263,0.98828) (0.01579,0.98943) (0.01895,0.99030) (0.02368,0.99128) (0.03000,0.99225) (0.03789,0.99315) (0.04579,0.99382) (0.05447,0.99423) (0.06632,0.99479) (0.07737,0.99570) (0.08763,0.99645) (0.10895,0.99743) (0.12237,0.99777) (0.15474,0.99817) (0.18395,0.99820) (0.31193,0.99830)};
        \addlegendentry{CNN, $L_\mathrm{b}=64$}
        \addplot[rocblue, rocdash] coordinates {(0.00354,0.73770) (0.00395,0.89518) (0.00474,0.93176) (0.00553,0.94978) (0.00632,0.96128) (0.00711,0.96717) (0.00789,0.97074) (0.00947,0.97574) (0.01105,0.98052) (0.01184,0.98249) (0.01263,0.98397) (0.01342,0.98495) (0.01500,0.98644) (0.01658,0.98754) (0.01895,0.98887) (0.02132,0.98993) (0.02526,0.99132) (0.03000,0.99276) (0.03316,0.99341) (0.03789,0.99394) (0.04579,0.99452) (0.05684,0.99509) (0.07184,0.99592) (0.08684,0.99675) (0.10500,0.99723) (0.13974,0.99782) (0.16974,0.99810) (0.31193,0.99877)};
        \addlegendentry{GLRT, $L_\mathrm{b}=64$}
        \addplot[rocpurple] coordinates {(0.00065,0.73770) (0.00079,0.89897) (0.00158,0.93043) (0.00237,0.94170) (0.00316,0.94857) (0.00395,0.95266) (0.00474,0.95524) (0.00553,0.95744) (0.00711,0.96147) (0.00868,0.96629) (0.00947,0.96787) (0.01026,0.96892) (0.01184,0.97050) (0.01500,0.97302) (0.01658,0.97404) (0.01816,0.97479) (0.02053,0.97559) (0.02605,0.97722) (0.03474,0.98004) (0.04184,0.98202) (0.04579,0.98286) (0.05132,0.98366) (0.06000,0.98459) (0.08211,0.98663) (0.10579,0.98848) (0.12474,0.99000) (0.13658,0.99132) (0.15158,0.99298) (0.15947,0.99358) (0.16579,0.99381) (0.19500,0.99409) (0.28579,0.99453) (0.31193,0.99461)};
        \addlegendentry{CNN, $L_\mathrm{b}=32$}
        \addplot[rocpurple, rocdash] coordinates {(0.00662,0.73770) (0.00711,0.81397) (0.00789,0.83985) (0.00947,0.87088) (0.01026,0.88617) (0.01105,0.90000) (0.01184,0.91182) (0.01263,0.92108) (0.01342,0.92802) (0.01421,0.93422) (0.01500,0.93969) (0.01579,0.94437) (0.01658,0.94818) (0.01737,0.95107) (0.01816,0.95309) (0.01895,0.95479) (0.01974,0.95625) (0.02132,0.95863) (0.02289,0.96058) (0.02763,0.96597) (0.03000,0.96828) (0.03158,0.96960) (0.03316,0.97071) (0.03553,0.97204) (0.03868,0.97350) (0.04263,0.97500) (0.05447,0.97907) (0.06000,0.98066) (0.06474,0.98172) (0.07184,0.98296) (0.08132,0.98428) (0.11526,0.98857) (0.12316,0.98915) (0.13737,0.98987) (0.16421,0.99087) (0.22342,0.99299) (0.29526,0.99524) (0.31193,0.99572)};
        \addlegendentry{GLRT, $L_\mathrm{b}=32$}
        \addplot[rocgold] coordinates {(0.00074,0.73770) (0.00079,0.78431) (0.00158,0.86187) (0.00237,0.88585) (0.00316,0.89736) (0.00395,0.90450) (0.00474,0.90983) (0.00553,0.91460) (0.00632,0.91887) (0.00711,0.92245) (0.00789,0.92554) (0.00947,0.93086) (0.01105,0.93568) (0.01184,0.93784) (0.01263,0.93975) (0.01421,0.94298) (0.01579,0.94574) (0.01737,0.94809) (0.01974,0.95103) (0.02211,0.95352) (0.02447,0.95557) (0.02684,0.95728) (0.03000,0.95922) (0.03553,0.96220) (0.04026,0.96454) (0.04421,0.96624) (0.04816,0.96762) (0.05368,0.96922) (0.06237,0.97136) (0.07342,0.97389) (0.10263,0.98035) (0.11368,0.98275) (0.12158,0.98418) (0.13974,0.98706) (0.15079,0.98855) (0.16026,0.98957) (0.16974,0.99029) (0.19105,0.99161) (0.20842,0.99241) (0.22342,0.99283) (0.24395,0.99306) (0.29289,0.99337) (0.31193,0.99341)};
        \addlegendentry{CNN, $L_\mathrm{b}=16$}
        \addplot[rocgold, rocdash] coordinates {(0.01671,0.73770) (0.01737,0.79651) (0.01816,0.81695) (0.01895,0.83005) (0.01974,0.84095) (0.02053,0.84996) (0.02132,0.85740) (0.02211,0.86360) (0.02289,0.86887) (0.02447,0.87790) (0.02605,0.88591) (0.02763,0.89286) (0.02921,0.89878) (0.03079,0.90373) (0.03158,0.90586) (0.03237,0.90776) (0.03316,0.90944) (0.03474,0.91228) (0.03632,0.91469) (0.03789,0.91677) (0.04026,0.91944) (0.04658,0.92604) (0.04974,0.92988) (0.05526,0.93673) (0.05763,0.93938) (0.06000,0.94168) (0.06158,0.94296) (0.06316,0.94399) (0.06632,0.94552) (0.07026,0.94712) (0.07500,0.94867) (0.08132,0.95039) (0.10579,0.95661) (0.12789,0.96182) (0.15158,0.96798) (0.15789,0.96921) (0.16342,0.96999) (0.16895,0.97046) (0.18000,0.97112) (0.19500,0.97171) (0.22184,0.97274) (0.23368,0.97351) (0.24553,0.97458) (0.25816,0.97599) (0.27158,0.97778) (0.28579,0.97997) (0.30079,0.98255) (0.31193,0.98462)};
        \addlegendentry{GLRT, $L_\mathrm{b}=16$}
      \end{axis}
    \end{tikzpicture}}
  \caption{Detection performance for different $L_\mathrm{b}$, $\lambda = 10^{-3}$. The GLRT benchmark employs GLRT-based start detection and correlation-based tail detection.}
  \label{fig:ROC}
\end{figure}

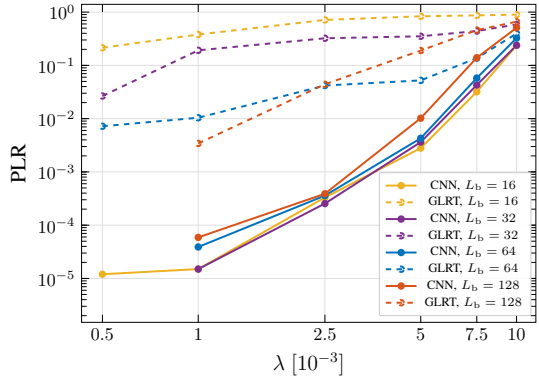
\begin{figure}
    \centering
    \resizebox{0.85\columnwidth}{!}{%
        \definecolor{mycolor1}{rgb}{0.929,0.694,0.125} 
\definecolor{mycolor2}{rgb}{0.494,0.184,0.556} 
\definecolor{mycolor3}{rgb}{0.000,0.447,0.741} 
\definecolor{mycolor4}{rgb}{0.850,0.325,0.098} 

\begin{tikzpicture}
\begin{axis}[
    width=\columnwidth,
    height=0.7\columnwidth,
    scale only axis,
    xmode=log,
    ymode=log,
    xmin=4.3e-4,
    xmax=1.15e-2,
    ymin=1.5e-6,
    ymax=1.4,
    xlabel={$\lambda\;[10^{-3}]$},
    ylabel={PLR},
    xtick={5e-4,1e-3,2.5e-3,5e-3,7.5e-3,1e-2},
    xticklabels={$0.5$,$1$,$2.5$,$5$,$7.5$,$10$},
    tick label style={font=\normalsize},
    label style={font=\large},
    major grid style={draw=gray!25},
    xmajorgrids,
    ymajorgrids,
    axis line style={black},
    tick style={black},
    legend style={
        at={(0.99,0.02)},
        anchor=south east,
        draw=gray!40,
        rounded corners=1pt,
        fill=white,
        fill opacity=0.88,
        text opacity=1,
        font=\scriptsize,
        legend columns=1,
        row sep=-1pt,
        column sep=2pt
    },
]


\addplot[
    color=mycolor1,
    line width=1.1pt,
    solid,
    mark=*,
    mark size=1.5pt
]
coordinates {
    (5e-4, 1.2e-5)
    (1e-3, 1.5e-5)
    (2.5e-3, 3.31e-4)
    (5e-3, 2.8e-3)
    (7.5e-3, 3.19e-2)
    (1e-2, 2.359e-1)
};
\addlegendentry{CNN, $L_{\mathrm b}=16$}

\addplot[
    color=mycolor1,
    line width=1.1pt,
    dashed,
    mark=o,
    mark size=1.5pt
]
coordinates {
    (5e-4, 2.14e-1)
    (1e-3, 3.8e-1)
    (2.5e-3, 7.13e-1)
    (5e-3, 8.36e-1)
    (7.5e-3, 8.68e-1)
    (1e-2, 8.97e-1)
};
\addlegendentry{GLRT, $L_{\mathrm b}=16$}


\addplot[
    color=mycolor2,
    line width=1.1pt,
    solid,
    mark=*,
    mark size=1.5pt
]
coordinates {
    (1e-3, 1.5e-5)
    (2.5e-3, 2.55e-4)
    (5e-3, 3.63e-3)
    (7.5e-3, 4.28e-2)
    (1e-2, 2.39e-1)
};
\addlegendentry{CNN, $L_{\mathrm b}=32$}

\addplot[
    color=mycolor2,
    line width=1.1pt,
    dashed,
    mark=o,
    mark size=1.5pt
]
coordinates {
    (5e-4, 2.66e-2)
    (1e-3, 1.92e-1)
    (2.5e-3, 3.23e-1)
    (5e-3, 3.52e-1)
    (7.5e-3, 4.41e-1)
    (1e-2, 5.86e-1)
};
\addlegendentry{GLRT, $L_{\mathrm b}=32$}


\addplot[
    color=mycolor3,
    line width=1.1pt,
    solid,
    mark=*,
    mark size=1.5pt
]
coordinates {
    (1e-3, 3.9e-5)
    (2.5e-3, 3.59e-4)
    (5e-3, 4.28e-3)
    (7.5e-3, 5.78e-2)
    (1e-2, 3.25e-1)
};
\addlegendentry{CNN, $L_{\mathrm b}=64$}

\addplot[
    color=mycolor3,
    line width=1.1pt,
    dashed,
    mark=o,
    mark size=1.5pt
]
coordinates {
    (5e-4, 7.2e-3)
    (1e-3, 1.04e-2)
    (2.5e-3, 4.19e-2)
    (5e-3, 5.23e-2)
    (7.5e-3, 1.37e-1)
    (1e-2, 3.96e-1)
};
\addlegendentry{GLRT, $L_{\mathrm b}=64$}


\addplot[
    color=mycolor4,
    line width=1.1pt,
    solid,
    mark=*,
    mark size=1.5pt
]
coordinates {
    (1e-3, 5.9e-5)
    (2.5e-3, 3.89e-4)
    (5e-3, 1.02e-2)
    (7.5e-3, 1.39e-1)
    (1e-2, 5.08e-1)
};
\addlegendentry{CNN, $L_{\mathrm b}=128$}

\addplot[
    color=mycolor4,
    line width=1.1pt,
    dashed,
    mark=o,
    mark size=1.5pt
]
coordinates {
    (1e-3, 3.43e-3)
    (2.5e-3, 4.42e-2)
    (5e-3, 1.9e-1)
    (7.5e-3, 4.6e-1)
    (1e-2, 6.6e-1)
};
\addlegendentry{GLRT, $L_{\mathrm b}=128$}

\end{axis}
\end{tikzpicture}
    }
 \caption{\ac{PLR} versus activation rate for different $L_{\mathrm{b}}$ and receiver architectures, over an observation interval of $20\,000$ symbols. The GLRT benchmark employs GLRT-based start detection and correlation-based tail detection.}
    \label{fig:plr}
\end{figure}

For the end-to-end results, the decision thresholds are selected on the validation data to target false-alarm rates of $F_A=0.05$ for start detection and $F_B=0.01$ for tail detection, and are then kept fixed throughout the simulations. For the \ac{PLR} and effective payload-efficiency curves, each operating point is simulated until one of two stopping criteria is met: either the number of observed packet errors exceeds $3\,000$, or the number of simulated observation intervals exceeds $25\,000$. Fig.~\ref{fig:plr} reports the end-to-end \ac{PLR} measured over observation intervals of $T=20\,000$ symbols. Commands are generated so that all their replicas lie within the observation interval, with no preceding traffic. These restrictions reduce interference near the edges of the observation interval, making the reported \ac{PLR} values optimistic relative to steady-state operation. Since the virtual-frame duration depends on $L_{\mathrm b}$, the edge effect also varies across boundary configurations. For the proposed method, the \ac{UPER} remained below 0.002\% at all operating points. With the proposed \ac{CNN}, $L_{\mathrm b}=16$ and $L_{\mathrm b}=32$ achieve comparable \ac{PLR} at $\lambda=10^{-3}$ and $\lambda=2.5\cdot10^{-3}$, whereas $L_{\mathrm b}=16$ gives the lowest \ac{PLR} from $\lambda=5\cdot10^{-3}$ onward. Shorter boundaries make detection harder, but the \ac{CNN} limits the resulting losses, allowing the reduced collision exposure to improve performance as traffic increases. For the \ac{GLRT} with the truncated sequences, detection losses at $L_{\mathrm b}=16$ and $32$ outweigh the benefit of shorter transmissions, and $L_{\mathrm b}=64$ gives the lowest \ac{PLR} at moderate and high loads.  

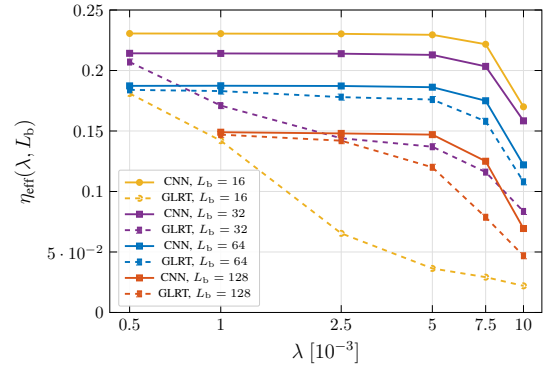
\begin{figure}
    \centering
    \resizebox{0.85\columnwidth}{!}{%
        \definecolor{mycolor1}{rgb}{0.929,0.694,0.125} 
\definecolor{mycolor2}{rgb}{0.494,0.184,0.556} 
\definecolor{mycolor3}{rgb}{0.000,0.447,0.741} 
\definecolor{mycolor4}{rgb}{0.850,0.325,0.098} 

\begin{tikzpicture}
\begin{axis}[
    width=\columnwidth,
    height=0.7\columnwidth,
    scale only axis,
    xmode=log,
    ymode=linear,
    xmin=4.3e-4,
    xmax=1.15e-2,
    ymin=1e-5,
    ymax=2.5e-1,
    xlabel={$\lambda\;[10^{-3}]$},
    ylabel={$\eta_{\text{eff}}(\lambda,L_\mathrm b)$},
    xtick={5e-4,1e-3,2.5e-3,5e-3,7.5e-3,1e-2},
    xticklabels={$0.5$,$1$,$2.5$,$5$,$7.5$,$10$},
    tick label style={font=\normalsize},
    label style={font=\large},
    major grid style={draw=gray!25},
    xmajorgrids,
    ymajorgrids,
    axis line style={black},
    tick style={black},
    legend style={
        at={(0.02,0.02)},
        anchor=south west,
        draw=gray!40,
        rounded corners=1pt,
        fill=white,
        fill opacity=0.88,
        text opacity=1,
        font=\scriptsize,
        legend columns=1,
        row sep=-1pt
    },
]


\addplot[
    color=mycolor1,
    line width=1.1pt,
    solid,
    mark=*,
    mark size=1.5pt
]
coordinates {
    (5e-4, 2.306e-1)
    (1e-3, 2.305e-1)
    (2.5e-3, 2.303e-1)
    (5e-3, 2.295e-1)
    (7.5e-3, 2.217e-1)
    (1e-2, 1.7e-1)
};
\addlegendentry{CNN, $L_{\mathrm b}=16$}

\addplot[
    color=mycolor1,
    line width=1.1pt,
    dashed,
    mark=o,
    mark size=1.5pt
]
coordinates {
    (5e-4, 1.81e-1)
    (1e-3, 1.42e-1)
    (2.5e-3, 6.54e-2)
    (5e-3, 3.64e-2)
    (7.5e-3, 2.92e-2)
    (1e-2, 2.2e-2)
};
\addlegendentry{GLRT, $L_{\mathrm b}=16$}


\addplot[
    color=mycolor2,
    line width=1.1pt,
    solid,
    mark=square*,
    mark size=1.5pt
]
coordinates {
    (5e-4, 2.142e-1)
    (1e-3, 2.141e-1)
    (2.5e-3, 2.139e-1)
    (5e-3, 2.129e-1)
    (7.5e-3, 2.034e-1)
    (1e-2, 1.584e-1)
};
\addlegendentry{CNN, $L_{\mathrm b}=32$}

\addplot[
    color=mycolor2,
    line width=1.1pt,
    dashed,
    mark=square*,
    mark size=1.5pt
]
coordinates {
    (5e-4, 2.07e-1)
    (1e-3, 1.71e-1)
    (2.5e-3, 1.44e-1)
    (5e-3, 1.37e-1)
    (7.5e-3, 1.16e-1)
    (1e-2, 8.35e-2)
};
\addlegendentry{GLRT, $L_{\mathrm b}=32$}


\addplot[
    color=mycolor3,
    line width=1.1pt,
    solid,
    mark=square*,
    mark size=1.5pt
]
coordinates {
    (5e-4, 1.873e-1)
    (1e-3, 1.874e-1)
    (2.5e-3, 1.872e-1)
    (5e-3, 1.862e-1)
    (7.5e-3, 1.75e-1)
    (1e-2, 1.22e-1)
};
\addlegendentry{CNN, $L_{\mathrm b}=64$}

\addplot[
    color=mycolor3,
    line width=1.1pt,
    dashed,
    mark=square*,
    mark size=1.5pt
]
coordinates {
    (5e-4,  1.84 e-1)
    (1e-3, 1.83e-1)
    (2.5e-3, 1.78e-1)
    (5e-3, 1.76e-1)
    (7.5e-3, 1.58e-1)
    (1e-2, 1.08e-1)
};
\addlegendentry{GLRT, $L_{\mathrm b}=64$}


\addplot[
    color=mycolor4,
    line width=1.1pt,
    solid,
    mark=square*,
    mark size=1.5pt
]
coordinates {
    (1e-3, 1.49e-1)
    (2.5e-3, 1.48e-1)
    (5e-3, 1.47e-1)
    (7.5e-3, 1.25e-1)
    (1e-2, 6.94e-2)
};
\addlegendentry{CNN, $L_{\mathrm b}=128$}

\addplot[
    color=mycolor4,
    line width=1.1pt,
    dashed,
    mark=square*,
    mark size=1.5pt
]
coordinates {
    (1e-3, 1.47e-1)
    (2.5e-3, 1.42e-1)
    (5e-3, 1.2e-1)
    (7.5e-3, 7.87e-2)
    (1e-2, 4.69 e-2)
};
\addlegendentry{GLRT, $L_{\mathrm b}=128$}

\end{axis}
\end{tikzpicture}
    }
    \caption{Effective payload efficiency versus activation rate for different boundary lengths and receiver architectures. The GLRT benchmark employs GLRT-based start detection and correlation-based tail detection.}
    \label{fig:goodput}
\end{figure}

Fig.~\ref{fig:goodput} reports the effective payload efficiency in \eqref{eq:effective_efficiency}, combining framing overhead and unsuccessful packet recovery. For the proposed \ac{CNN}, $L_{\mathrm b}=16$ provides the highest efficiency at every simulated activation rate. At low load, its value approaches the loss-free efficiency, reflecting the substantial overhead reduction obtained by shortening the boundaries. As the traffic increases, its \ac{PLR} grows, but the combination of lower framing overhead and reduced collision exposure remains sufficient to preserve its advantage over the longer configurations. With the truncated sequences, the \ac{GLRT} cannot fully exploit very short boundaries: $L_{\mathrm b}=64$ provides the highest efficiency for $\lambda\geq10^{-3}$, since the detection losses at $L_{\mathrm b}=16$ and $32$ outweigh their overhead advantage.

\section{Conclusion}
\label{sec:con}

Short-boundary signaling removes the architectural constraint between framing and coding in asynchronous grant-free \ac{C2C} communications, turning the boundary length into an independent system-design parameter.  The analysis shows that shorter boundaries improve both framing efficiency and the probability of avoiding temporal overlap, while the idealized \ac{MSE} analysis quantifies the benefit of decision-directed refinement. The proposed  \ac{CNN} adapts its waveform branch to the shorter boundaries while retaining the complete decoder \ac{LLR} trajectory for tail detection. For the considered setting, the results show that the boundary length should not be inherited from the channel code: the proposed receiver makes boundary shortening viable, with the shortest tested boundary providing the highest effective payload efficiency throughout the investigated traffic range.

\bibliographystyle{IEEEtran}
\bibliography{Files/biblio}

\end{document}